\documentclass[aps,pre,reprint,amsmath,superscriptaddress]{revtex4-1}
\usepackage{amsmath}
\usepackage[latin1]{inputenc}
\usepackage{graphicx}
\usepackage[]{hyperref}
\usepackage{bm}

\usepackage{epsfig}

\usepackage{color}
\usepackage{xcolor}
\hypersetup{
    colorlinks,
    linkcolor={red!50!black},
    citecolor={blue!50!black},
    urlcolor={blue!80!black}
}

\def\minus{
\scalebox{0.4}[1.0]{\( - \)}
}

\begin{document}
\title{Topologically protected colloidal transport above a square magnetic lattice}
\begin{abstract}
We theoretically study the motion of magnetic colloidal particles above a magnetic pattern and compare the predictions with Brownian dynamics simulations. The pattern
consists of alternating square domains of positive and negative magnetization. The colloidal motion is driven by periodic modulation loops of
an external magnetic field. There exist loops that induce topologically protected colloidal transport between two different unit cells of the pattern. 
The transport is very robust against internal and external perturbations. Theory and simulations are in perfect agreement. 
Our theory is applicable to other systems with the same symmetry.
\end{abstract}
\author{Daniel de las Heras$^a$}
\author{Johannes Loehr$^b$}
\author{Michael Loenne$^c$}
\author{Thomas M. Fischer$^b$}
\email{thomas.fischer@uni-bayreuth.de}
\affiliation{
Theoretische Physik$^{\;a}$, 
Experimentalphysik$^{\;b}$, and
Mathematik$^{\;c}$,
Institutes of Physics and Mathematics, Universit\"at Bayreuth, 95440 Bayreuth, Germany.}
\date{\today}

\maketitle

\section{Introduction}

Controlling the transport of colloidal particles is a requisite in several applications such as lab-on-a-chip devices~\cite{C1LC20683D},
drug delivery with colloidal carriers~\cite{Kataoka2001113,muller1991colloidal}, and computation with colloids~\cite{C4SM00796D}. 

Techniques to control the motion of colloids include the use of gradient fields~\cite{0953-8984-20-40-404215}, thermal ratchets~\cite{matthias2003asymmetric,PhysRevLett.91.060602,B910427E}, liquid
crystal-based solvents~\cite{PhysRevE.64.031711,Turiv1351}, and active particles~\cite{PhysRevLett.105.268302}. Colloidal particles are usually polydisperse in e.g. size, mass, etc. 
Therefore, the transport of a collection of colloids using the above techniques results always in a dispersion of the motion. One 
can avoid this by using optical tweezers~\cite{grier2003revolution} but at the expenses of having to move the colloids on a one-by-one basis. 

Topological protection is a promising tool to overcome these problems. If the dynamics depends
only on a topological invariant it is possible to have total control over the colloidal motion, independently
of the intrinsic characteristics of the particles. Recently, we have studied the motion of magnetic
colloids above a hexagonal magnetic pattern~\cite{N6}. The system is driven by an external magnetic field.
The positions of the colloids above the pattern are given by the minima of the magnetic potential 
which has contributions from the static field of the pattern and the time dependent external field. 
The set of stationary points of the potential form a surface in the full phase space whose topological
properties fully determine the colloidal motion. There exist transport modes that are topologically protected and
therefore extremely robust against perturbations.  

The topology of the stationary surface, and hence the topologically protected transport modes, are unique for each type of lattice. 
Here, we theoretically study the transport of diamagnetic colloidal particles above a square magnetic lattice, and compare the results with computer simulations.

\section{Theory}

\begin{figure}
\vspace{0.2cm}
\includegraphics[width=0.5\columnwidth]{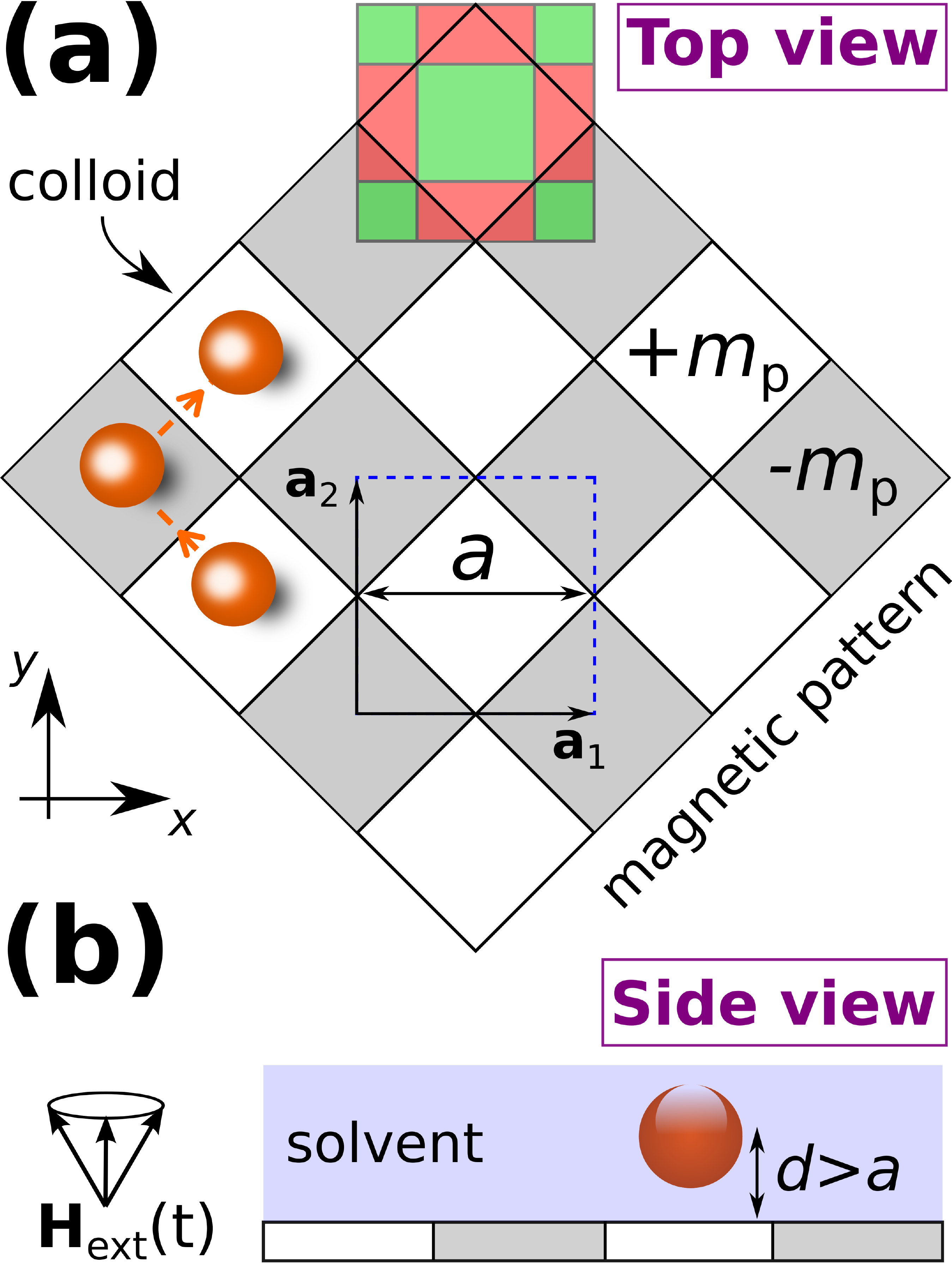}
\caption{Schematic top- (a) and side- (b) views of the system.
The pattern is a periodic lattice of squares with diagonal length $a$ and alternating positive and negative magnetization perpendicular
to the film, $+m_{\text p}$ and
$-m_{\text p}$, respectively. 
A time-dependent external magnetic field $\mathbf{H}_{\text{ext}}(t)$ drives the system.
The diamagnetic colloids (orange spheres) are located at a distance $d>a$ above the pattern.
A unit cell, square of side-length $a$, is highlighted with a blue-dashed line in (a). Another unit cell (top of panel a) is coloured indicating 
the allowed (green) and forbidden (red) regions for the colloids.
} 
\label{fig1}
\end{figure}

The colloids move in a plane at a distance $d>a$ above the pattern, with $a$ the side-length of the unit cell of the pattern, see 
Fig.~\ref{fig1}. A time-dependent external magnetic field $\mathbf{H}_{\text{ext}}(t)$ drives the system. The variation in time
of $\mathbf{H}_{\text{ext}}(t)$ is slow enough such that the colloidal particles can adiabatically follow the minima of the magnetic
potential at any time $t$. The magnetic potential is $V=-\chi_{\text{eff}}\mu_0\mathbf{H}\cdot{\mathbf{H}}$, where $\mathbf{H}$ is the total magnetic field with contributions from the square pattern and the external
potential, $\chi_{\text eff}<0$ is the effective magnetic susceptibility of the diamagnets in the solvent, and ${\mu_0}$ is the vacuum permeability. 

\begin{figure*}
\includegraphics[width=0.75\textwidth]{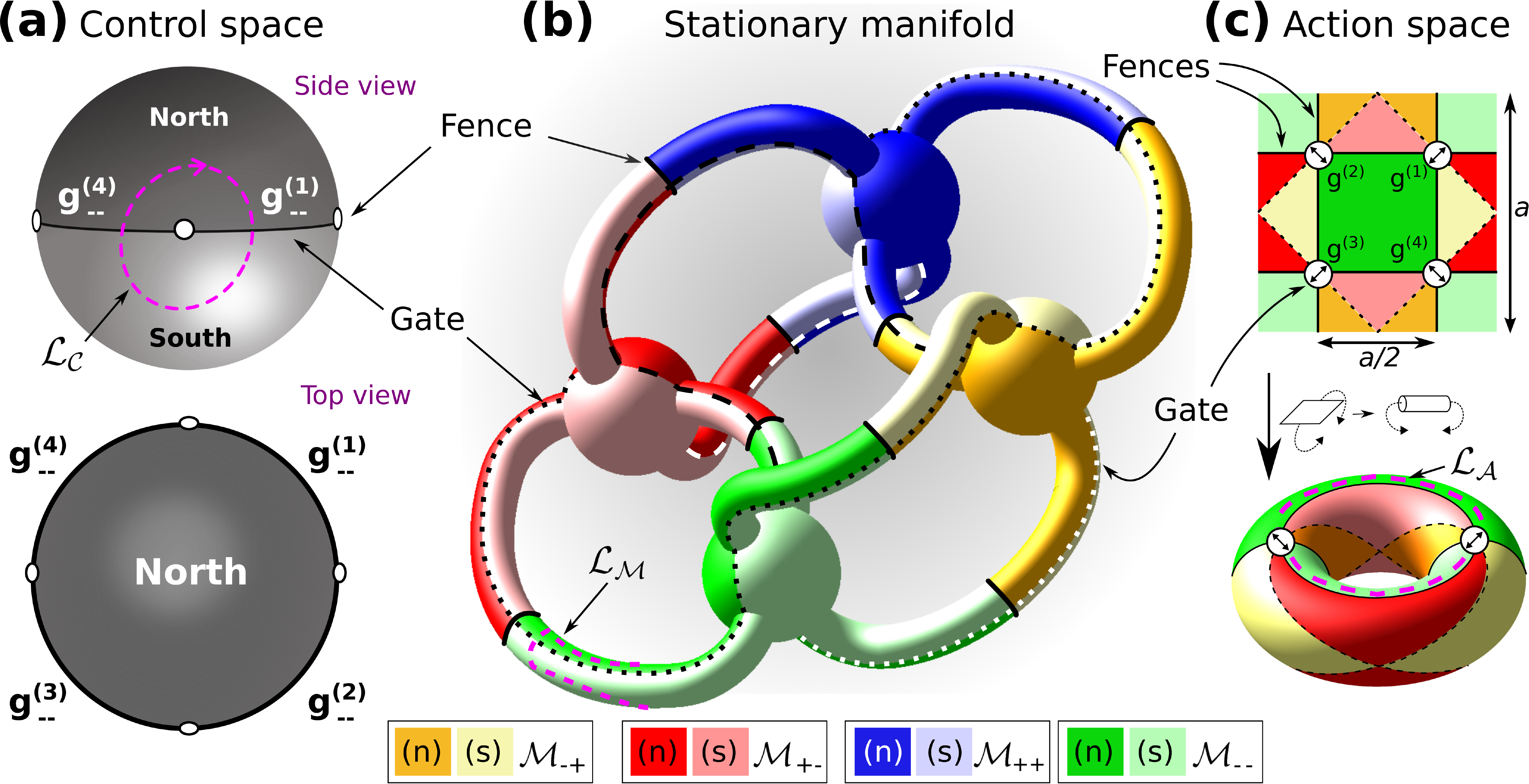}
\caption{Control space ${\cal C}$ (a), the stationary surface ${\cal M}$ (b), and action space ${\cal A}$ (c). Each color in ${\cal M}$ (b) represents a bijective
area, as indicated (dark colors for northern areas and soft colors for southern areas). The solid lines are fences and the dotted and dashed lines are gates.
In ${\cal C}$ (a) the solid lines on the equator are the segments of minima of the gates and the  empty circles the fences. In ${\cal A}$ (c) the fences are represented by solid lines and the gates
by circles with arrows indicating the possible transport directions. The color of ${\cal A}$ is given by the projection of half of ${\cal M}$ into ${\cal A}$. The violet dashed
line in (a) is a control loop, ${\cal L_C}=(g_1,\bar g_4)$, that crosses two gates and induces colloidal transport.
The preimage loop in ${\cal M}_{--}$ is indicated by ${\cal L_M}$ and the corresponding loop in ${\cal A}$ by ${\cal L_A}$.} 
\label{fig2}
\end{figure*}

$\mathbf{H}$ can be expressed as a Fourier series with Fourier modes that decay exponentially with $z$. Hence, 
at high elevations, $z>a$, the potential is well approximated by $V\propto\mathbf{H}_{\text{ext}}(t)\cdot\mathbf{H}_{\text p}(\mathbf{x}_{\cal A})$, where
\begin{equation}
\mathbf{H}_{\text{p}}(\mathbf{x}_{\cal A})\propto
\sum_{i=1}^4\left(\begin{array}{c}
q_{i,x}\sin(\mathbf{q}_{i}\cdot \mathbf{x}_{\cal A})\\
q_{i,y}\sin(\mathbf{q}_{i}\cdot \mathbf{x}_{\cal A})\\
q\cos(\mathbf{q}_{i}\cdot \mathbf{x}_{\cal A})
\end{array}\right),
\label{eq1}
\end{equation}
is, up to a multiplicative constant, the contribution from the magnetic pattern. Here,
\begin{equation}
\mathbf{q}_i=\frac{2\pi}{a}\left(\begin{array}{c}\sin(2\pi i/4)\\-\cos(2\pi i/4)\end{array}\right),\text{  }i=1,..,4,
\end{equation}
are the reciprocal lattice vectors of the second Brillouin zone with $q=2\pi/a$ their common magnitude. $\mathbf{x}_{\cal A}=x_1\mathbf{a}_1+x_2\mathbf{a}_2$
with $\mathbf{a}_i$ the basic lattice vectors of the square pattern (see Fig.~\ref{fig1}), are the coordinates in {\it action space} ${\cal A}$, i.e,
the plane above the pattern in which the colloids move. We vary $\mathbf{H}_{\text{ext}}(t)$ on the surface of a sphere,
\begin{equation}
\mathbf{H}_{\text{ext}}=H_{\text{ext}}(\cos\phi_t\sin\theta_t,\sin\phi_t\sin\theta_t,\cos\theta_t).
\end{equation}
The set $(\theta_t,\phi_t)$ define our {\it control space}, ${\cal C}$, see Fig.~\ref{fig2}a. 
We measure $\theta_t$ with respect to the $z$ axis and $\phi_t$ with respect to $\mathbf{a}_1$. The system is driven with periodic closed loops of $\mathbf{H}_{\text{ext}}(t)$.
There exist special loops that induce transport between different unit cells, i.e., when $\mathbf{H}_{\text{ext}}$ returns to its initial position the particle is in a 
different unit cell. 

To understand the motion we need to look at the full phase space, i.e, the product space ${\cal C}\otimes{\cal A}$,
with states given by $(\mathbf{H}_{\text{ext}}(t),\mathbf{x}_{\cal A})$. The stationary 
points satisfy $\nabla_{\cal A}V=0$, with $\nabla_{\cal A}$ the gradient in ${\cal A}$. The 
set of all stationary points is a two dimensional manifold in ${\cal C}\otimes{\cal A}$ that we call the {\it stationary manifold}, ${\cal M}$, see Fig.~\ref{fig2}b. 

The correspondence between ${\cal M}$ and ${\cal C}$ is not bijective.
Each direction of the external field is a point in ${\cal C}$. For each point in ${\cal C}$ (with the exception of four special points that we discuss later)
there are four points (preimages) in ${\cal M}$, the
solutions of $\nabla_{\cal A}V=0$. Two solutions are saddle points of $V$, one is a maximum, and the other one is a minimum. In ${\cal A}$ the four points form a square of side $a/2$. 

The correspondence between ${\cal M}$ and ${\cal A}$ is also not bijective. Consider the unit vectors 
$\mathbf{\hat e}_i(\mathbf{x}_A)=\partial_i\mathbf{H}_{\text{p}}/|\partial_i\mathbf{H}_{\text{p}}|$, $i=1,2$. Then, a point $\mathbf{x}_{A}$ in ${\cal A}$ is stationary if the external
field points in a direction perpendicular to both $\mathbf{\hat e}_1$ and $\mathbf{\hat e}_2$, i.e.,
\begin{equation}
\mathbf{H}_{\text{ext}}^{(s)}(\mathbf{x}_{\cal A})=\pm H_{\text{ext}}\frac{\mathbf{\hat e}_1\times\mathbf{\hat e}_2}{|\mathbf{\hat e}_1\times\mathbf{\hat e}_2|}.
\label{hsta}
\end{equation}
The subscript $(s)$ stands for stationary. That is,
each point in ${\cal A}$ has two preimages $(\mathbf{H}_{\text{ext}}^{(s)},\mathbf{x}_{\cal A})$ in ${\cal M}$ (except for special points that we describe later). 

Consider now the matrix of the second derivatives of $V$ evaluated at the stationary field
\begin{equation}
\nabla_{\cal A}\nabla_{\cal A}V|_{{\mathbf H}_{{\text{ext}}}^{\text{(s)}}}=\left( {\begin{array}{cc} 
 {\mathbf H}_{{\text{ext}}}^{\text{(s)}}\cdot\partial_1\partial_1{\mathbf H_{\text{p}}} & 
 {\mathbf H}_{{\text{ext}}}^{\text{(s)}}\cdot\partial_1\partial_2{\mathbf H_{\text{p}}} \cr
 {\mathbf H}_{{\text{ext}}}^{\text{(s)}}\cdot\partial_2\partial_1{\mathbf H_{\text{p}}} & 
 {\mathbf H}_{{\text{ext}}}^{\text{(s)}}\cdot\partial_2\partial_2{\mathbf H_{\text{p}}} \cr
\end{array} } \right)\label{Hessian},
\end{equation}
which is diagonal since the mixed derivatives vanish, cf.~\eqref{eq1}.
The stationary manifold ${\cal M}$ is the union of submanifolds ${\cal M}_{\alpha\beta}$,
where $\alpha$ ($\beta$) is the opposite sign of the eigenvalue of \eqref{Hessian} with eigenvector 
pointing in  the $\mathbf{a}_1$ ($\mathbf{a}_2$) direction. 
That is, 
${\cal M}={\cal M}_{++}\cup{\cal M}_{+-}\cup{\cal M}_{-+}\cup{\cal M}_{--}$.
Hence the stable trajectories for the colloids reside in ${\cal M}_{\minus\minus}$ (minima of $V$). ${\cal M}_{++}$ are maxima of $V$, and both ${\cal M}_{+-}$ 
and ${\cal M}_{-+}$ are saddle points. All the submanifolds are topologically equivalent
since each point in ${\cal C}$ has one preimage in each of the submanifolds. 

The submanifolds share common borders in ${\cal M}$ that we call the {\it fences}.
Any two submanifolds with one common sign of one of the eigenvalues are glued together in ${\cal M}$ through two fences.
At the fences one eigenvalue changes its sign, i.e., the determinant of~\eqref{Hessian} vanishes. For example,
${\cal M}_{++}$ and ${\cal M}_{-+}$ share two fences. At both fences the eigenvalue of the eigenvector pointing 
along the $\mathbf{a_1}$ direction changes it sign. The stationary field, Eq.~\eqref{hsta}, points along $+\mathbf{a_2}$ 
in one fence and along $-\mathbf{a_2}$ in the other fence. Hence, in ${\cal M}$ we have four submanifolds, and each one is double-joined
to other two submanifolds. In other words, ${\cal M}$ is a genus $5$ surface, see Fig.~\ref{fig2}b. 

Solving 
$||\nabla_{\cal A}\nabla_{\cal A}V||=0$ we can see the fences in action and control space. In ${\cal C}$ the fences are four equispaced points along the equator,
corresponding to external fields pointing in $\pm\mathbf{a}_1$ and $\pm\mathbf{a}_2$, see Fig.~\ref{fig2}a. The fences divide action space in a 
square lattice (length $a/2$) of alternating allowed and forbidden regions, see Fig.~\ref{fig2}c and Fig.~\ref{fig1}a. 
Using periodic boundary conditions ${\cal A}$ is a torus.
The allowed regions are areas of minima of $V$ (projection of the submanifold ${\cal M}_{--}$ into ${\cal A}$). In the forbidden areas all the stationary points are saddle points. 
As we have seen, a point in ${\cal A}$ can be made stationary with two opposite external fields. Therefore ${\cal M}_{++}$ and ${\cal M}_{--}$ are projected into the same regions in ${\cal A}$. 
In other words, if there is a minimum of the potential in a given point in ${\cal A}$ we can turn it into a maximum by just pointing the external field in the opposite direction.
${\cal M}_{+-}$ and ${\cal M}_{-+}$ are also projected into the same areas in ${\cal A}$. In Fig.~\ref{fig2}c we show the projection of half of ${\cal M}$ into ${\cal A}$ (the half
that contains all points closer to ${\cal M}_{\minus\minus}$ than to ${\cal M}_{++}$)
such that each area has a unique meaning. That is, the projection of this half of ${\cal M}$ into ${\cal A}$ is bijective. 
 
The fences cross in ${\cal A}$ at points that we call the {\it gates} since they connect two allowed regions in ${\cal A}$.
There are four gates $g^{(i)}$, $i=1,..,4$, see Fig.~\ref{fig2}c. The gates play a vital role for the colloidal motion.
To find the gates in ${\cal C}$ we note that the fences do
not cross in ${\cal M}$ but they do cross in ${\cal A}$. Hence, $\mathbf{H}_{\text{ext}}^{(s)}$ cannot be unique at the gates in ${\cal A}$
(crossing points between fences in ${\cal A}$). The only possibility is that $\mathbf{\hat e}_1$ is parallel to $\mathbf{\hat e}_2$, see Eq.~\eqref{hsta},
at the gates. Therefore, as $\mathbf{H}_{\text{ext}}^{(s)}\perp\mathbf{\hat e}_1,\mathbf{\hat e}_2$,
the gates in ${\cal C}$ are great circles. For the present square lattice the gates in ${\cal C}$ are located on the equator. Each gate is divided in four segments, $g^{(i)}_{\alpha\beta}$ where
$\alpha,\beta=\pm$ are again the opposite signs of the eigenvalues of \eqref{Hessian}.
Although all gates in ${\cal C}$ are in the equator, they are rotated such that the union of four segments with
identical signs of the eigenvalues form a full equator, see Fig.~\ref{fig2}a. 

The gates split ${\cal C}$ in two parts, the south $(s)$ and the north $(n)$, see Fig.~\ref{fig2}a. They also split each submanifold of ${\cal M}$ in two parts
${\cal M}_{\alpha\beta}={\cal M}_{\alpha\beta}^{(n)}\cup{\cal M}_{\alpha\beta}^{(s)}$, see Fig.~\ref{fig2}b. This splitting is very convenient since the resulting
regions ${\cal M}_{\alpha\beta}^{(\nu)}$ with $\nu=n,s$ are simply connected bijective areas. That is, there are no holes in ${\cal M}_{\alpha\beta}^{(\nu)}$ and 
the correspondences between ${\cal M}_{\alpha\beta}^{(\nu)}$ and the other spaces (${\cal C}$ and ${\cal A}$) are unique.

\section{Results}
\begin{figure*}
\includegraphics[width=0.60\textwidth]{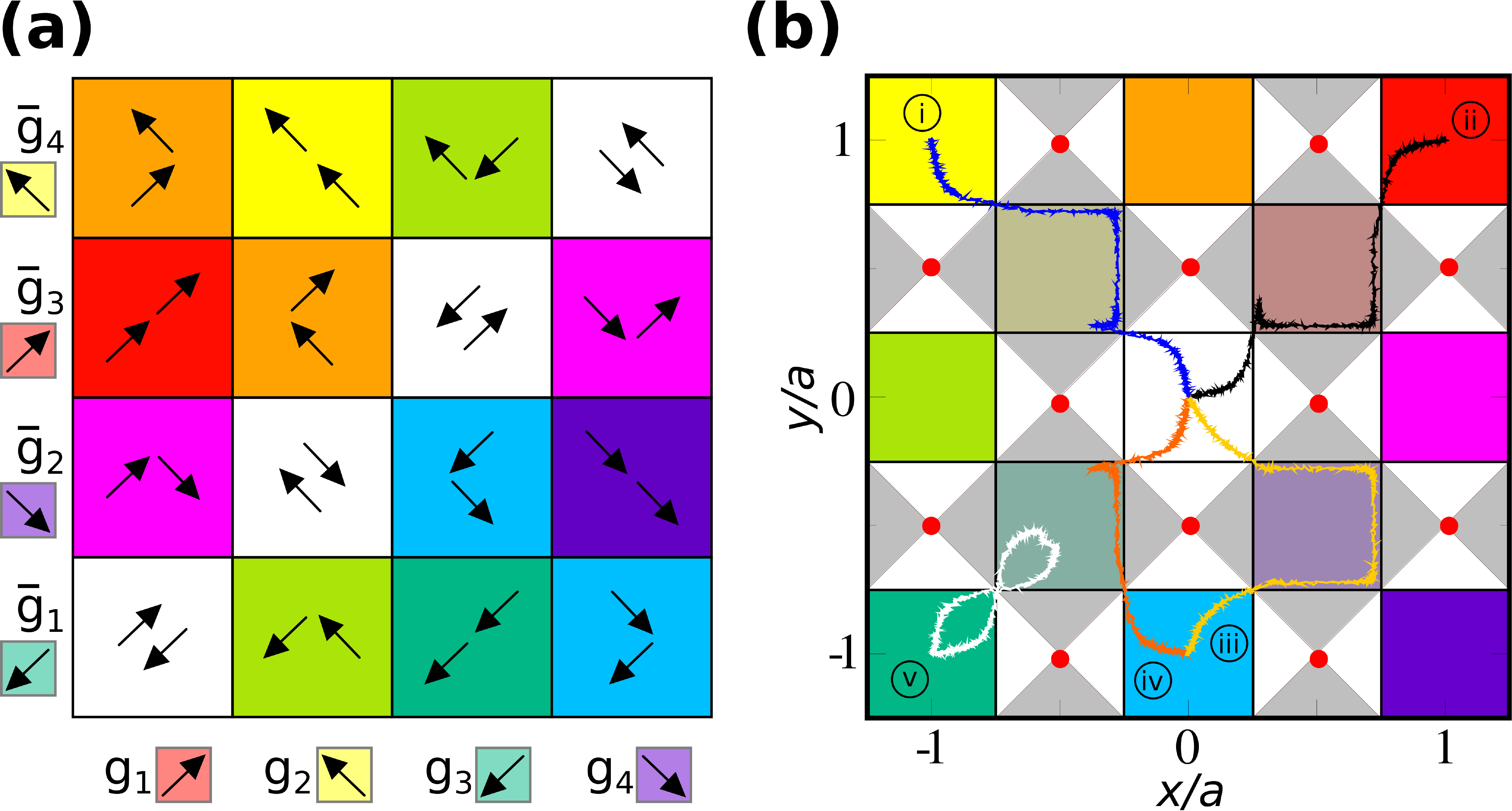}
\caption{(a) Phase diagram of the colloidal motion in the plane $g_i-\bar g_j$ for the fundamental modulation loops in control space  ${\cal L_C}=(g_{i},\bar g_{j})$. 
The loop starts in the north of ${\cal C}$ then goes to the south using the gate segment $g^{(i)}_{--}$ and returns to the south trough the segment $g^{(j)}_{--}$.
Each color represents a transport direction. The arrows indicate which gates are crossed and in which sense.
(b) Examples of the trajectories of the colloids in ${\cal A}$ according to BD simulations for the modulation loops: (i) ${\cal L_C}=(g_2,\bar g_4)$,
(ii) ${\cal L_C}=(g_1,\bar g_3)$, (iii) ${\cal L_C}=(g_4,\bar g_1)$, (iv) ${\cal L_C}=(g_3,\bar g_2)$, and (v) ${\cal L_C}=(g_1,\bar g_1)$.
The solid lines are the fences in ${\cal A}$. The forbidden regions are marked with a middle red circle.
The allowed regions are coloured according to the phase diagram in (a). We show four trajectories, (i) to (iv), corresponding to
loops that induce colloidal transport (the initial position of the colloids is the allowed region centered at the origin), and one trajectory corresponding
to a topologically trivial control loop (v) that does not induce transport (the initial position of the colloid is the allowed region
centered at $x/a=-1$ and $y/a=-1$). The magnetic pattern is also represented using white and grey regions.}  
\label{fig3}
\end{figure*}

We are now in a position to understand the colloidal motion. Let ${\cal L_C}$ be a closed modulation loop of the external field in $\cal C$.
${\cal L_C}$ has four preimage loops in ${\cal M}$, one in each submanifold ${\cal M}_{\alpha\beta}$.
Only the loop lying in ${\cal M}_{\minus\minus}$ is populated with colloids. This populated loop can be then projected into ${\cal A}$ where we can read the actual trajectory of the colloids. 
Loops ${\cal L_C}$ that induce colloidal transport from one unit cell to another in ${\cal A}$ are only those that cross at least two different gates in control space,
which is equivalent to enclosing at least one fence in ${\cal C}$. When ${\cal L_C}$ crosses the segment $g^{(i)}_{\minus\minus}$ in ${\cal C}$, the corresponding loop
that transport the colloids in ${\cal A}$ also crosses the gate $g^{(i)}$. Each gate in ${\cal C}$ can be crossed from the north to the south or from the south to the north,
which in ${\cal A}$ results in opposite senses. Let ${\cal L_C}=(g_{i},\bar g_{j})$ be a loop of the external field that starts on the north of ${\cal C}$, then goes to the south of ${\cal C}$
crossing the segment of minima of the potential of the gate $i$ ($g^{(i)}_{\minus\minus})$ and returns to the initial point in the north of ${\cal C}$ using the segment of minima of the gate $j$. 
An example of such a loop is represented in Fig.~\ref{fig2}a. The phase diagram of the colloidal
motion in the $g_{i}-\bar g_{j}$ plane is depicted in Fig.~\ref{fig3}a. It has been obtained (i) theoretically by translating loops in ${\cal C}$ into loops in ${\cal A}$ using the stationary
surface ${\cal M}$ and (ii) with standard Brownian dynamics simulations. Details of the simulations
are provided in the Appendix. The agreement between theory and simulations is perfect.

Loops that cross the same gate twice, i.e, ${\cal L_C}=(g_{i},\bar g_{i})$, do not induce transport
between different unit cells (the initial and the final positions are the same). Loops that cross
different gates induce transport between nearest or second nearest unit cells. There are two possible
routes for each of the nearest unit cells (see e.g. ${\cal L_C}=(g_{1},\bar g_{4})$
and $(g_{2},\bar g_{3})$) and only one in the case of second nearest unit cells (e.g., 
${\cal L_C}=(g_{1},\bar g_{3})$. In Fig.~\ref{fig3}b we show Brownian dynamics trajectories for selected modulation loops.  

\begin{figure*}
\includegraphics[width=0.60\textwidth]{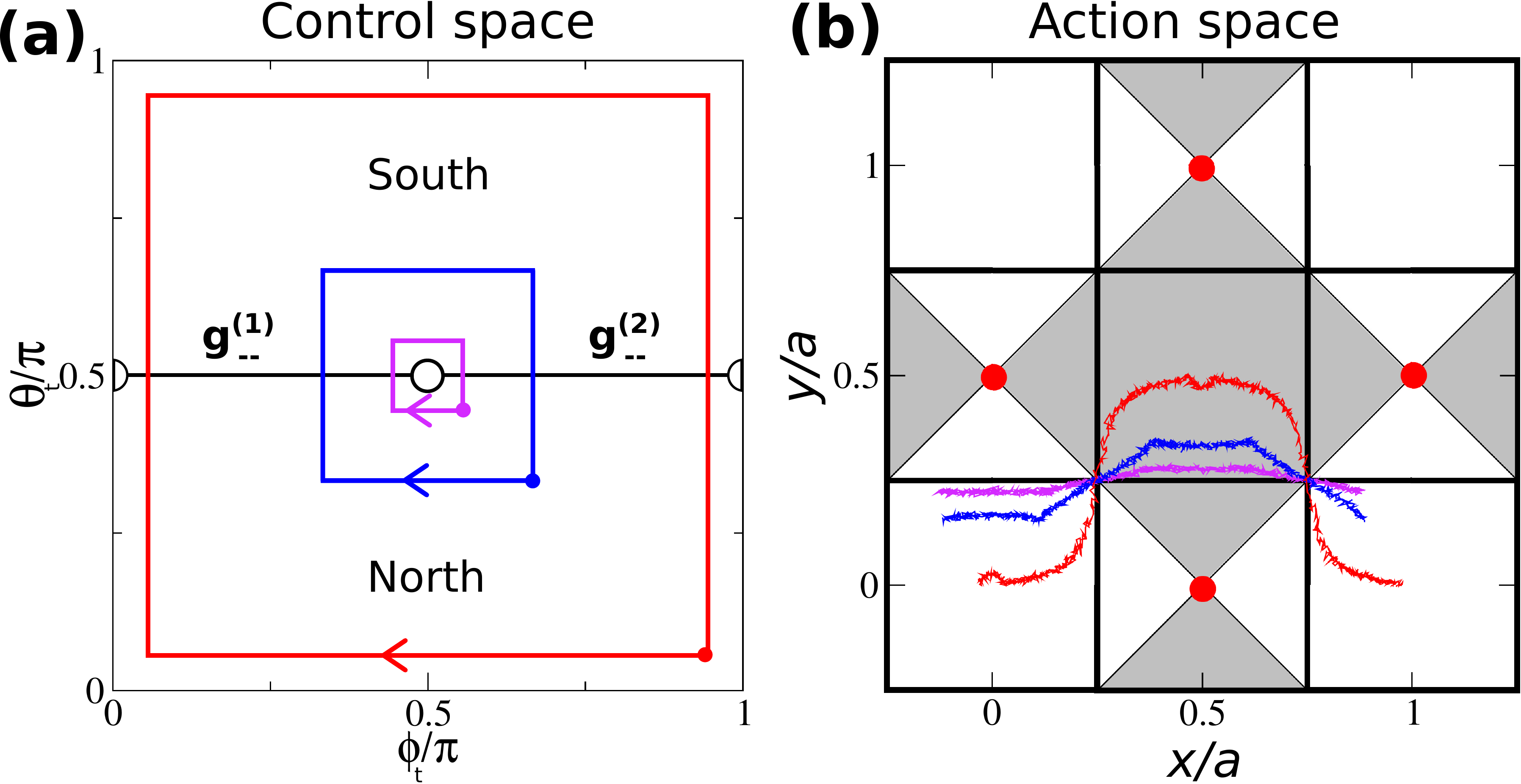}
\caption{(a) Modulations loops in control space of type ${\cal L_C}=(g_1,\bar g_2)$. The direction and the starting point of the loops are indicated by arrows and
filled circles, respectively. The empty circles are the fences in control space and the horizontal black line are the gates as indicated. (b) Trajectories in
action space corresponding to the loops showed in (a). The trajectories are coloured according to the color of the loops in (a). 
The white and grey areas indicate the magnetic pattern. The squares are the allowed and forbidden areas of action
space. The forbidden  areas are highlighted with a red circle in the middle. The initial position of the colloids is the allowed area centered at the origin.}
\label{fig4}
\end{figure*}

\begin{figure*}
\includegraphics[width=0.75\textwidth]{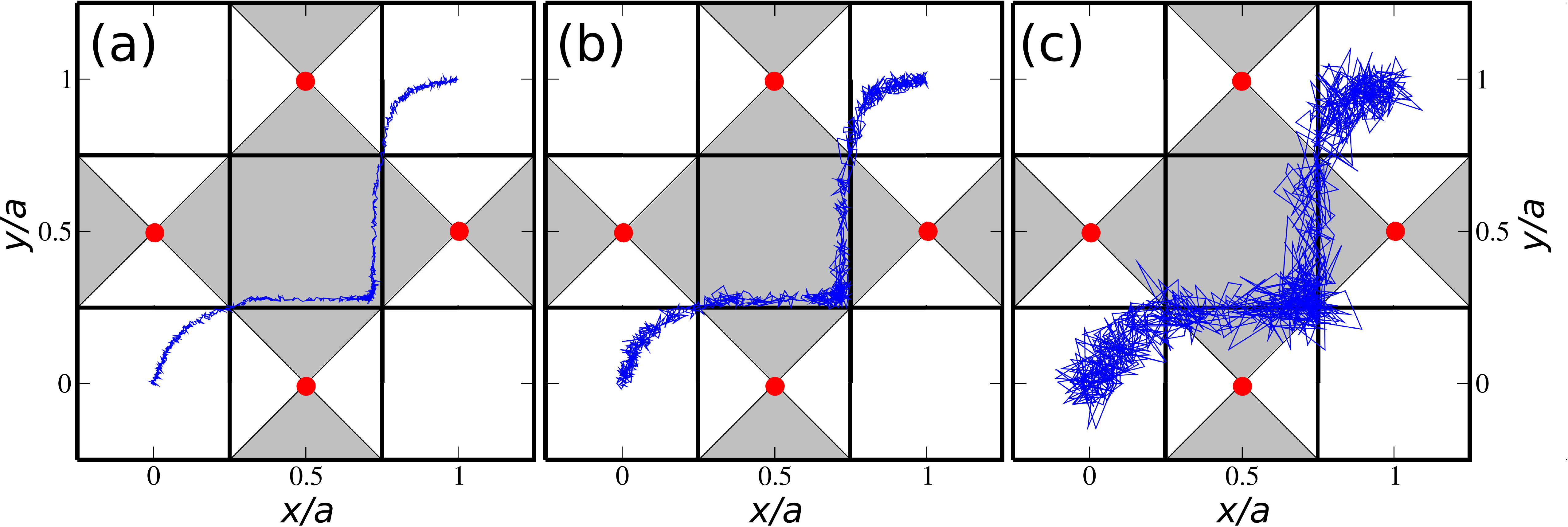}
\caption{Trajectories in action space of a diamagnetic colloid under a control loop ${\cal L_C}=(g_1,\bar g_3)$ for different values of the scaled temperature $k_{\text B}T/\epsilon=0.01$ (a), $0.1$ (b),
and $1.0$ (c). Here $k_\text{B}$ is the Boltzmann constant, and $\epsilon$ sets the unit of energy of the magnetic potential $V$. The white and grey areas indicate the magnetic pattern. The squares are the allowed and forbidden areas of action
space. The forbidden  areas are highlighted with a red circle in the middle. The initial position of the colloids is the allowed area centered at the origin.}
\label{fig5}
\end{figure*}

The colloidal transport is very robust against internal and external perturbations.
The shape of ${\cal L_C}$, for example, is completely irrelevant.
Only the gates that ${\cal L_C}$ crosses are important. 
In Fig.~\ref{fig4} we show the trajectories in action space 
for three modulation loops that cross the same two gates, ${\cal L_C}=(g_1,\bar g_2)$, yet following different paths. 
The trajectories in ${\cal A}$ differ but the starting and ending allowed regions are the same. 
The motion is also robust against changes in the speed of the modulation, the thermal noise, 
and properties of the colloidal particles such as size, mass, effective susceptibility, etc (see an example in Fig.~\ref{fig5}). 
Therefore we can transport in a dispersion-free and
precise way a collection of particles with a broad distribution of masses, sizes, etc. 

The reason behind this robustness is that the transport direction depends only on
a topological invariant, and hence it is topologically protected. 
For each loop in ${\cal M}$ we can define a set of $10$ winding numbers, two for each hole of ${\cal M}$.
$S_{\cal M}$, the set of winding numbers of the loop in ${\cal M}_{\minus\minus}$, is the topological invariant. 
In each of the regions of the phase diagram $S_{\cal M}$ does not vary. Alternatively we can define the topological invariant of loops in ${\cal A}$ and ${\cal C}$. 
The loop that lies in ${\cal M_{--}}$ is projected into a loop in ${\cal A}$ and ${\cal C}$. Since ${\cal M_{\minus\minus}}$ is topologically equivalent to control 
space without the fences, ${\cal C'}$,
the correspondence between loops in ${\cal M}_{\minus\minus}$ and ${\cal C'}$ is bijective.
$S_{\cal C}$, the set of winding numbers of loops around the fences in ${\cal C'}$ induce corresponding winding numbers
of loops around the torus in action space ($S_{\cal A}=\{w_1,w_2\}$ with $w_i=0,\pm1$) via the
loops in ${\cal M}_{\minus\minus}$.
Each of the eight non-zero values of $S_{\cal A}$ corresponds to a type of transport in ${\cal A}$.
$S_{\cal A}$ and $S_{\cal C}$ are also topological invariants, they
remain unchanged for each type of transport, i.e, in each region of the phase
diagram of Fig.~\ref{fig3}a. 

\begin{figure}
\includegraphics[width=0.69\columnwidth]{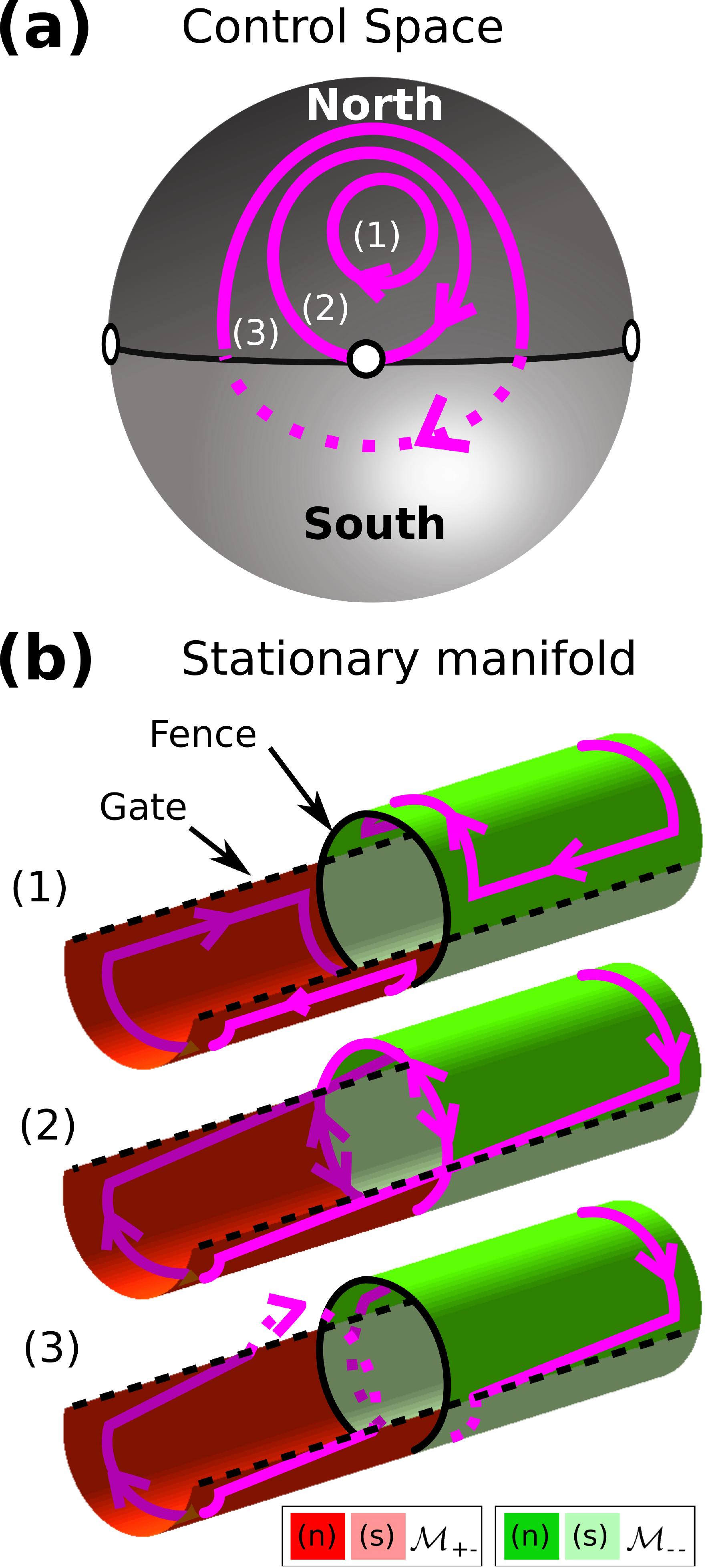}
\caption{Control space ${\cal C}$ (a) and part of the stationary manifold ${\cal M}$ (b). Three modulation loops ${\cal L_C}$ in control space and their corresponding
loops in ${\cal M_{--}}$ and ${\cal M_{+-}}$ are indicated by violet lines. The loops are represented with a solid (dashed)
line in the north (south) of control space and the stationary manifold. The arrows indicate the direction of the loops.}
\label{fig6}
\end{figure}

How is it possible to change the direction of transport if it is topologically protected? There are always operations 
that break the topological protection. This is precisely what happens at the interface between two transport directions in the phase diagram, cf. Fig.~\ref{fig3}a.
At the interfaces between two different transport modes
the topological protection is lost allowing for a change
in the transport mode. This occurs for modulation loops that cross at
least one of the fences in control space. In Fig.~\ref{fig6} we show
an example of this process. The loop labeled as (1) lies entirely on the north of ${\cal C}$.
That is, it does not cross gates and hence does not induce transport between different unit
cells. The corresponding loops in ${\cal M}$ lie on the northern areas
of ${\cal M}$. There is one loop in each of the submanifolds of ${\cal M}$. 
Fig.~\ref{fig6} shows only the loops in ${\cal M}_{--}$ and ${\cal M}_{+-}$. When the loop in ${\cal C}$
touches one of the fences (see loop (2) in Fig.~\ref{fig6}) the loops in ${\cal M}_{--}$ and ${\cal M}_{+-}$
join at the fence (the loops in ${\cal M}_{++}$ and 
${\cal M}_{-+}$ also join at a different fence).
At this point the colloids, which follow the loop in ${\cal M}_{--}$, have two alternative paths:
(i) a loop that resides entirely in the north of ${\cal M}_{--}$ and (ii) a loop that lies in both the north
and the south of ${\cal M}_{\minus\minus}$ and hence induce colloidal transport between different cells.
The motion is not topologically protected in the sense that two different trajectories
are possible. Next, we expand the loop in ${\cal C}$ such that it encloses one fence in ${\cal C}$ and hence crosses two 
gates, see loop (3) in Fig.~\ref{fig6}. In ${\cal M}$ the loops in ${\cal M}_{\minus\minus}$ and ${\cal M}_{+-}$ are now disjoined and
have interchanged a segment at the fence. The result is two loops that no longer reside
in the northern areas of ${\cal M}$. The loop in ${\cal M}_{\minus\minus}$ winds around the holes of ${\cal M}$ inducing colloidal transport.
The direction of transport has changed with respect to the initial loop (1).

Due to the thermal noise in Brownian dynamics simulations the particles fluctuate around the minima of the potential,
exploring the neighborhood of ${\cal M}_{\minus\minus}$ in ${\cal C}\otimes{\cal A}$. Hence, modulation loops in control space that do not cross a
fence, but pass close enough to it, might also be topologically unprotected, leading to two differing transport modes in ${\cal A}$. How close
the control loop has to be to the fence in order to be deprotected depends on the magnitude of the thermal noise. The thermal noise effectively
expand the fences in ${\cal C}$ into the surrounding areas, and broaden the topological transition in ${\cal A}$.

\section{Discussion}

We have explained the motion of diamagnetic colloids for which the effective susceptibility is negative. Paramagnetic colloids have a positive effective susceptibility,
and hence will follow the maxima of $V$. 
The minima and the maxima of $V$ always comove in ${\cal A}$ separated by $\mathbf{r}=(a/2,a/2)$. Therefore, paramagnetic colloids perform
the same motion as diamagnets but displaced by $\mathbf{r}$.

From a experimental view point, it is possible to use magnetic bubble lattices~\cite{bubble} or lithographic patterns~\cite{lito} to generate the pattern. 
Possible methods to levitate the colloids above the pattern consist on using a ferrofluid solvent~\cite{N6} and the deposition of a polymer layer~\cite{PhysRevLett.112.048302}
on the magnetic pattern.

The colloidal transport is fully determined by the topology of the manifold ${\cal M}$, which is unique for each type of magnetic pattern. For example,
the stationary manifold of a hexagonal pattern is a genus $7$ surface~\cite{N6}. There, the modulation loops in ${\cal C}$ that induce transport of colloids must 
cross the fences in ${\cal C}$, which are lines instead of points as in the present study. As a result, transport modes of hexagonal and square patterns
are completely different. In both, hexagonal and square lattices, the topological invariant in ${\cal A}$ is the set of two winding numbers
around the hole in ${\cal A}$. This is just a consequence of the dimension of ${\cal A}$. Control space ${\cal C}$ neither contains all the information. For example,
in square lattices the transition between transport modes occurs for those loops that cross a fence. However, in hexagonal lattices, a fence crossing loop in ${\cal C}$ 
is a necessary but not sufficient condition to change the transport mode. What fully determines the transport modes is the stationary manifold
${\cal M}$ (the topology, the fences, and how ${\cal M}$ is projected into ${\cal A}$ and ${\cal C}$). In ${\cal M}$ the topological invariant is the set of winding numbers around the holes, which is very different
in square (${\cal M}$ has genus $5$) and hexagonal (${\cal M}$ has genus $7$) lattices.

The topologically protected transport modes we have shown here can be understood as bulk modes sustained (driven) by an external field. The transport
occurs in the bulk of the periodic system.  Other forms of topologically protected motion occur at the edges of a periodic system, such as e.g., the motion of 
 electrons in topological insulators~\cite{RevModPhys.82.3045}, mechanical solitons~\cite{kane2014topological,chen2014nonlinear,paulose2015topological}, phonons~\cite{huber2016topological},
and photons~\cite{rechtsman2013photonic,lu2016topological} among others. There, a perturbation populates
an edge state that cannot scatter into the bulk due to the topology of the system. Our theory is transferable to other systems with the same symmetry. Hence, topological bulk states might exist in e.g.
excitons in superlattices~\cite{PhysRevB.40.1357,kim2011dynamical}, tight-binding models~\cite{PhysRevLett.106.236803}, and cold atoms in optical lattices~\cite{PhysRevA.90.013609}. 
Topologically protected edge states might also occur at the borders of finite magnetic lattices. Their topological properties might be substantially different from
those of bulk states. How the edge states in our particle system compare to other edge states in wave systems is a very interesting subject for future studies.

In wave systems, such as e.g. topological insulators, the topology of the band structure is characterized by the Chern numbers of the bands. Each Chern number
can be computed as an integral over the Berry curvature of the band~\cite{dirac}. In our particle system we describe the topological protection in terms of the stationary manifold. Both 
descriptions are probably equivalent in some form. 

\begin{acknowledgments}
This publication was funded by the German Research Foundation (DFG) and the University of Bayreuth in the funding programme Open Access Publishing.
\end{acknowledgments}

\appendix
\section{Brownian Dynamics Simulations}
We use Brownian Dynamics to simulate the motion of a diamagnetic colloid above the pattern. The
coordinates in action space are ${\bf x}_{\cal A}$, and the equation of motion is given
by
\begin{equation}
\xi\frac{d{\bf x_{\cal A}}(t)}{dt}=-\nabla_{\cal A}V({\bf x}_{\cal A},{\bf H}_{\text{ext}}(t))+\bm{\eta}(t),\nonumber
\end{equation}
where $t$ is the time, $\xi$ is the friction coefficient, and $\bm\eta$ is a Gaussian random force with a variance given by
the fluctuation-dissipation theorem. The magnetic potential $V$ has contributions from the external field ${\bf H}_{\text{ext}}$
and the magnetic pattern (see the main text).

The equation of motion is integrated in time with a standard Euler algorithm:
\begin{equation}
{\bf x_{\cal A}}(t+\Delta t)={\bf x_{\cal A}}(t)-\nabla_{\cal A}V\Delta t+\bm{\delta}{\bf r},
\end{equation}
where $\Delta t$ is the time step, and $\bm{\delta}{\bf r}$ is a random displacement sampled from a gaussian distribution with standard deviation $\sqrt{2\Delta tk_{\text{B}}T/\xi}$.
Here $k_{\text{B}}$ is the Boltzmann constant, and $T$ is absolute temperature. Before starting the modulation loop in $\cal{C}$ we first equilibrate the system
by running $10^4$ time steps such that the colloids find the minimum of the magnetic potential at $t=0$.


\begin{thebibliography}{28}%
\makeatletter
\providecommand \@ifxundefined [1]{%
 \@ifx{#1\undefined}
}%
\providecommand \@ifnum [1]{%
 \ifnum #1\expandafter \@firstoftwo
 \else \expandafter \@secondoftwo
 \fi
}%
\providecommand \@ifx [1]{%
 \ifx #1\expandafter \@firstoftwo
 \else \expandafter \@secondoftwo
 \fi
}%
\providecommand \natexlab [1]{#1}%
\providecommand \enquote  [1]{``#1''}%
\providecommand \bibnamefont  [1]{#1}%
\providecommand \bibfnamefont [1]{#1}%
\providecommand \citenamefont [1]{#1}%
\providecommand \href@noop [0]{\@secondoftwo}%
\providecommand \href [0]{\begingroup \@sanitize@url \@href}%
\providecommand \@href[1]{\@@startlink{#1}\@@href}%
\providecommand \@@href[1]{\endgroup#1\@@endlink}%
\providecommand \@sanitize@url [0]{\catcode `\\12\catcode `\$12\catcode
  `\&12\catcode `\#12\catcode `\^12\catcode `\_12\catcode `\%12\relax}%
\providecommand \@@startlink[1]{}%
\providecommand \@@endlink[0]{}%
\providecommand \url  [0]{\begingroup\@sanitize@url \@url }%
\providecommand \@url [1]{\endgroup\@href {#1}{\urlprefix }}%
\providecommand \urlprefix  [0]{URL }%
\providecommand \Eprint [0]{\href }%
\providecommand \doibase [0]{http://dx.doi.org/}%
\providecommand \selectlanguage [0]{\@gobble}%
\providecommand \bibinfo  [0]{\@secondoftwo}%
\providecommand \bibfield  [0]{\@secondoftwo}%
\providecommand \translation [1]{[#1]}%
\providecommand \BibitemOpen [0]{}%
\providecommand \bibitemStop [0]{}%
\providecommand \bibitemNoStop [0]{.\EOS\space}%
\providecommand \EOS [0]{\spacefactor3000\relax}%
\providecommand \BibitemShut  [1]{\csname bibitem#1\endcsname}%
\let\auto@bib@innerbib\@empty
\bibitem [{\citenamefont {Gao}\ \emph {et~al.}(2011)\citenamefont {Gao},
  \citenamefont {Tahir}, \citenamefont {Virgin},\ and\ \citenamefont
  {Yellen}}]{C1LC20683D}%
  \BibitemOpen
  \bibfield  {author} {\bibinfo {author} {\bibfnamefont {L.}~\bibnamefont
  {Gao}}, \bibinfo {author} {\bibfnamefont {M.~A.}\ \bibnamefont {Tahir}},
  \bibinfo {author} {\bibfnamefont {L.~N.}\ \bibnamefont {Virgin}}, \ and\
  \bibinfo {author} {\bibfnamefont {B.~B.}\ \bibnamefont {Yellen}},\ }\href
  {\doibase 10.1039/C1LC20683D} {\bibfield  {journal} {\bibinfo  {journal} {Lab
  Chip}\ }\textbf {\bibinfo {volume} {11}},\ \bibinfo {pages} {4214} (\bibinfo
  {year} {2011})}\BibitemShut {NoStop}%
\bibitem [{\citenamefont {Kataoka}\ \emph {et~al.}(2001)\citenamefont
  {Kataoka}, \citenamefont {Harada},\ and\ \citenamefont
  {Nagasaki}}]{Kataoka2001113}%
  \BibitemOpen
  \bibfield  {author} {\bibinfo {author} {\bibfnamefont {K.}~\bibnamefont
  {Kataoka}}, \bibinfo {author} {\bibfnamefont {A.}~\bibnamefont {Harada}}, \
  and\ \bibinfo {author} {\bibfnamefont {Y.}~\bibnamefont {Nagasaki}},\ }\href
  {\doibase http://dx.doi.org/10.1016/S0169-409X(00)00124-1} {\bibfield
  {journal} {\bibinfo  {journal} {Adv. Drug Deliv. Rev.}\ }\textbf {\bibinfo
  {volume} {47}},\ \bibinfo {pages} {113 } (\bibinfo {year}
  {2001})}\BibitemShut {NoStop}%
\bibitem [{\citenamefont {M{\"u}ller}(1991)}]{muller1991colloidal}%
  \BibitemOpen
  \bibfield  {author} {\bibinfo {author} {\bibfnamefont {R.~H.}\ \bibnamefont
  {M{\"u}ller}},\ }\href@noop {} {\emph {\bibinfo {title} {Colloidal carriers
  for controlled drug delivery and targeting: Modification, characterization
  and in vivo distribution}}}\ (\bibinfo  {publisher} {Taylor \& Francis},\
  \bibinfo {year} {1991})\BibitemShut {NoStop}%
\bibitem [{\citenamefont {Phillips}\ \emph {et~al.}(2014)\citenamefont
  {Phillips}, \citenamefont {Jankowski}, \citenamefont {Krishnatreya},
  \citenamefont {Edmond}, \citenamefont {Sacanna}, \citenamefont {Grier},
  \citenamefont {Pine},\ and\ \citenamefont {Glotzer}}]{C4SM00796D}%
  \BibitemOpen
  \bibfield  {author} {\bibinfo {author} {\bibfnamefont {C.~L.}\ \bibnamefont
  {Phillips}}, \bibinfo {author} {\bibfnamefont {E.}~\bibnamefont {Jankowski}},
  \bibinfo {author} {\bibfnamefont {B.~J.}\ \bibnamefont {Krishnatreya}},
  \bibinfo {author} {\bibfnamefont {K.~V.}\ \bibnamefont {Edmond}}, \bibinfo
  {author} {\bibfnamefont {S.}~\bibnamefont {Sacanna}}, \bibinfo {author}
  {\bibfnamefont {D.~G.}\ \bibnamefont {Grier}}, \bibinfo {author}
  {\bibfnamefont {D.~J.}\ \bibnamefont {Pine}}, \ and\ \bibinfo {author}
  {\bibfnamefont {S.~C.}\ \bibnamefont {Glotzer}},\ }\href {\doibase
  10.1039/C4SM00796D} {\bibfield  {journal} {\bibinfo  {journal} {Soft Matter}\
  }\textbf {\bibinfo {volume} {10}},\ \bibinfo {pages} {7468} (\bibinfo {year}
  {2014})}\BibitemShut {NoStop}%
\bibitem [{\citenamefont {Erbe}\ \emph {et~al.}(2008)\citenamefont {Erbe},
  \citenamefont {Zientara}, \citenamefont {Baraban}, \citenamefont {Kreidler},\
  and\ \citenamefont {Leiderer}}]{0953-8984-20-40-404215}%
  \BibitemOpen
  \bibfield  {author} {\bibinfo {author} {\bibfnamefont {A.}~\bibnamefont
  {Erbe}}, \bibinfo {author} {\bibfnamefont {M.}~\bibnamefont {Zientara}},
  \bibinfo {author} {\bibfnamefont {L.}~\bibnamefont {Baraban}}, \bibinfo
  {author} {\bibfnamefont {C.}~\bibnamefont {Kreidler}}, \ and\ \bibinfo
  {author} {\bibfnamefont {P.}~\bibnamefont {Leiderer}},\ }\href
  {http://stacks.iop.org/0953-8984/20/i=40/a=404215} {\bibfield  {journal}
  {\bibinfo  {journal} {J. Phys.: Condens. Matter}\ }\textbf {\bibinfo {volume}
  {20}},\ \bibinfo {pages} {404215} (\bibinfo {year} {2008})}\BibitemShut
  {NoStop}%
\bibitem [{\citenamefont {Matthias}\ and\ \citenamefont
  {M{\"u}ller}(2003)}]{matthias2003asymmetric}%
  \BibitemOpen
  \bibfield  {author} {\bibinfo {author} {\bibfnamefont {S.}~\bibnamefont
  {Matthias}}\ and\ \bibinfo {author} {\bibfnamefont {F.}~\bibnamefont
  {M{\"u}ller}},\ }\href {\doibase 10.1038/nature01736} {\bibfield  {journal}
  {\bibinfo  {journal} {Nature}\ }\textbf {\bibinfo {volume} {424}},\ \bibinfo
  {pages} {53} (\bibinfo {year} {2003})}\BibitemShut {NoStop}%
\bibitem [{\citenamefont {Engel}\ \emph {et~al.}(2003)\citenamefont {Engel},
  \citenamefont {M\"uller}, \citenamefont {Reimann},\ and\ \citenamefont
  {Jung}}]{PhysRevLett.91.060602}%
  \BibitemOpen
  \bibfield  {author} {\bibinfo {author} {\bibfnamefont {A.}~\bibnamefont
  {Engel}}, \bibinfo {author} {\bibfnamefont {H.~W.}\ \bibnamefont {M\"uller}},
  \bibinfo {author} {\bibfnamefont {P.}~\bibnamefont {Reimann}}, \ and\
  \bibinfo {author} {\bibfnamefont {A.}~\bibnamefont {Jung}},\ }\href {\doibase
  10.1103/PhysRevLett.91.060602} {\bibfield  {journal} {\bibinfo  {journal}
  {Phys. Rev. Lett.}\ }\textbf {\bibinfo {volume} {91}},\ \bibinfo {pages}
  {060602} (\bibinfo {year} {2003})}\BibitemShut {NoStop}%
\bibitem [{\citenamefont {Tierno}\ \emph {et~al.}(2009)\citenamefont {Tierno},
  \citenamefont {Sagues}, \citenamefont {Johansen},\ and\ \citenamefont
  {Fischer}}]{B910427E}%
  \BibitemOpen
  \bibfield  {author} {\bibinfo {author} {\bibfnamefont {P.}~\bibnamefont
  {Tierno}}, \bibinfo {author} {\bibfnamefont {F.}~\bibnamefont {Sagues}},
  \bibinfo {author} {\bibfnamefont {T.~H.}\ \bibnamefont {Johansen}}, \ and\
  \bibinfo {author} {\bibfnamefont {T.~M.}\ \bibnamefont {Fischer}},\ }\href
  {\doibase 10.1039/B910427E} {\bibfield  {journal} {\bibinfo  {journal} {Phys.
  Chem. Chem. Phys.}\ }\textbf {\bibinfo {volume} {11}},\ \bibinfo {pages}
  {9615} (\bibinfo {year} {2009})}\BibitemShut {NoStop}%
\bibitem [{\citenamefont {Stark}\ and\ \citenamefont
  {Ventzki}(2001)}]{PhysRevE.64.031711}%
  \BibitemOpen
  \bibfield  {author} {\bibinfo {author} {\bibfnamefont {H.}~\bibnamefont
  {Stark}}\ and\ \bibinfo {author} {\bibfnamefont {D.}~\bibnamefont
  {Ventzki}},\ }\href {\doibase 10.1103/PhysRevE.64.031711} {\bibfield
  {journal} {\bibinfo  {journal} {Phys. Rev. E}\ }\textbf {\bibinfo {volume}
  {64}},\ \bibinfo {pages} {031711} (\bibinfo {year} {2001})}\BibitemShut
  {NoStop}%
\bibitem [{\citenamefont {Turiv}\ \emph {et~al.}(2013)\citenamefont {Turiv},
  \citenamefont {Lazo}, \citenamefont {Brodin}, \citenamefont {Lev},
  \citenamefont {Reiffenrath}, \citenamefont {Nazarenko},\ and\ \citenamefont
  {Lavrentovich}}]{Turiv1351}%
  \BibitemOpen
  \bibfield  {author} {\bibinfo {author} {\bibfnamefont {T.}~\bibnamefont
  {Turiv}}, \bibinfo {author} {\bibfnamefont {I.}~\bibnamefont {Lazo}},
  \bibinfo {author} {\bibfnamefont {A.}~\bibnamefont {Brodin}}, \bibinfo
  {author} {\bibfnamefont {B.~I.}\ \bibnamefont {Lev}}, \bibinfo {author}
  {\bibfnamefont {V.}~\bibnamefont {Reiffenrath}}, \bibinfo {author}
  {\bibfnamefont {V.~G.}\ \bibnamefont {Nazarenko}}, \ and\ \bibinfo {author}
  {\bibfnamefont {O.~D.}\ \bibnamefont {Lavrentovich}},\ }\href {\doibase
  10.1126/science.1240591} {\bibfield  {journal} {\bibinfo  {journal}
  {Science}\ }\textbf {\bibinfo {volume} {342}},\ \bibinfo {pages} {1351}
  (\bibinfo {year} {2013})}\BibitemShut {NoStop}%
\bibitem [{\citenamefont {Jiang}\ \emph {et~al.}(2010)\citenamefont {Jiang},
  \citenamefont {Yoshinaga},\ and\ \citenamefont
  {Sano}}]{PhysRevLett.105.268302}%
  \BibitemOpen
  \bibfield  {author} {\bibinfo {author} {\bibfnamefont {H.-R.}\ \bibnamefont
  {Jiang}}, \bibinfo {author} {\bibfnamefont {N.}~\bibnamefont {Yoshinaga}}, \
  and\ \bibinfo {author} {\bibfnamefont {M.}~\bibnamefont {Sano}},\ }\href
  {\doibase 10.1103/PhysRevLett.105.268302} {\bibfield  {journal} {\bibinfo
  {journal} {Phys. Rev. Lett.}\ }\textbf {\bibinfo {volume} {105}},\ \bibinfo
  {pages} {268302} (\bibinfo {year} {2010})}\BibitemShut {NoStop}%
\bibitem [{\citenamefont {Grier}(2003)}]{grier2003revolution}%
  \BibitemOpen
  \bibfield  {author} {\bibinfo {author} {\bibfnamefont {D.~G.}\ \bibnamefont
  {Grier}},\ }\href {\doibase 10.1038/nature01935} {\bibfield  {journal}
  {\bibinfo  {journal} {Nature}\ }\textbf {\bibinfo {volume} {424}},\ \bibinfo
  {pages} {810} (\bibinfo {year} {2003})}\BibitemShut {NoStop}%
\bibitem [{\citenamefont {Loehr}\ \emph {et~al.}(2016)\citenamefont {Loehr},
  \citenamefont {Loenne}, \citenamefont {Ernst}, \citenamefont {de~las Heras},\
  and\ \citenamefont {Fischer}}]{N6}%
  \BibitemOpen
  \bibfield  {author} {\bibinfo {author} {\bibfnamefont {J.}~\bibnamefont
  {Loehr}}, \bibinfo {author} {\bibfnamefont {M.}~\bibnamefont {Loenne}},
  \bibinfo {author} {\bibfnamefont {A.}~\bibnamefont {Ernst}}, \bibinfo
  {author} {\bibfnamefont {D.}~\bibnamefont {de~las Heras}}, \ and\ \bibinfo
  {author} {\bibfnamefont {T.~M.}\ \bibnamefont {Fischer}},\ }\href {\doibase
  10.1038/ncomms11745} {\bibfield  {journal} {\bibinfo  {journal} {Nat.
  Commun.}\ }\textbf {\bibinfo {volume} {7}},\ \bibinfo {pages} {11745}
  (\bibinfo {year} {2016})}\BibitemShut {NoStop}%
\bibitem [{\citenamefont {Bobeck}\ \emph {et~al.}(1975)\citenamefont {Bobeck},
  \citenamefont {Bonyhard},\ and\ \citenamefont {Geusic}}]{bubble}%
  \BibitemOpen
  \bibfield  {author} {\bibinfo {author} {\bibfnamefont {A.~H.}\ \bibnamefont
  {Bobeck}}, \bibinfo {author} {\bibfnamefont {P.~I.}\ \bibnamefont
  {Bonyhard}}, \ and\ \bibinfo {author} {\bibfnamefont {J.~E.}\ \bibnamefont
  {Geusic}},\ }\href {\doibase 10.1109/PROC.1975.9912} {\bibfield  {journal}
  {\bibinfo  {journal} {Proc. IEEE}\ }\textbf {\bibinfo {volume} {63}},\
  \bibinfo {pages} {1176} (\bibinfo {year} {1975})}\BibitemShut {NoStop}%
\bibitem [{\citenamefont {Terris}\ and\ \citenamefont {Thomson}(2005)}]{lito}%
  \BibitemOpen
  \bibfield  {author} {\bibinfo {author} {\bibfnamefont {B.~D.}\ \bibnamefont
  {Terris}}\ and\ \bibinfo {author} {\bibfnamefont {T.}~\bibnamefont
  {Thomson}},\ }\href {http://stacks.iop.org/0022-3727/38/i=12/a=R01}
  {\bibfield  {journal} {\bibinfo  {journal} {J. Phys. D: Appl. Phys.}\
  }\textbf {\bibinfo {volume} {38}},\ \bibinfo {pages} {R199} (\bibinfo {year}
  {2005})}\BibitemShut {NoStop}%
\bibitem [{\citenamefont {Tierno}\ and\ \citenamefont
  {Fischer}(2014)}]{PhysRevLett.112.048302}%
  \BibitemOpen
  \bibfield  {author} {\bibinfo {author} {\bibfnamefont {P.}~\bibnamefont
  {Tierno}}\ and\ \bibinfo {author} {\bibfnamefont {T.~M.}\ \bibnamefont
  {Fischer}},\ }\href {\doibase 10.1103/PhysRevLett.112.048302} {\bibfield
  {journal} {\bibinfo  {journal} {Phys. Rev. Lett.}\ }\textbf {\bibinfo
  {volume} {112}},\ \bibinfo {pages} {048302} (\bibinfo {year}
  {2014})}\BibitemShut {NoStop}%
\bibitem [{\citenamefont {Hasan}\ and\ \citenamefont
  {Kane}(2010)}]{RevModPhys.82.3045}%
  \BibitemOpen
  \bibfield  {author} {\bibinfo {author} {\bibfnamefont {M.~Z.}\ \bibnamefont
  {Hasan}}\ and\ \bibinfo {author} {\bibfnamefont {C.~L.}\ \bibnamefont
  {Kane}},\ }\href {\doibase 10.1103/RevModPhys.82.3045} {\bibfield  {journal}
  {\bibinfo  {journal} {Rev. Mod. Phys.}\ }\textbf {\bibinfo {volume} {82}},\
  \bibinfo {pages} {3045} (\bibinfo {year} {2010})}\BibitemShut {NoStop}%
\bibitem [{\citenamefont {Kane}\ and\ \citenamefont
  {Lubensky}(2014)}]{kane2014topological}%
  \BibitemOpen
  \bibfield  {author} {\bibinfo {author} {\bibfnamefont {C.}~\bibnamefont
  {Kane}}\ and\ \bibinfo {author} {\bibfnamefont {T.}~\bibnamefont
  {Lubensky}},\ }\href {\doibase 10.1038/nphys2835} {\bibfield  {journal}
  {\bibinfo  {journal} {Nat. Phys.}\ }\textbf {\bibinfo {volume} {10}},\
  \bibinfo {pages} {39} (\bibinfo {year} {2014})}\BibitemShut {NoStop}%
\bibitem [{\citenamefont {Chen}\ \emph {et~al.}(2014)\citenamefont {Chen},
  \citenamefont {Upadhyaya},\ and\ \citenamefont
  {Vitelli}}]{chen2014nonlinear}%
  \BibitemOpen
  \bibfield  {author} {\bibinfo {author} {\bibfnamefont {B.~G.-g.}\
  \bibnamefont {Chen}}, \bibinfo {author} {\bibfnamefont {N.}~\bibnamefont
  {Upadhyaya}}, \ and\ \bibinfo {author} {\bibfnamefont {V.}~\bibnamefont
  {Vitelli}},\ }\href {\doibase 10.1073/pnas.1405969111} {\bibfield  {journal}
  {\bibinfo  {journal} {PNAS}\ }\textbf {\bibinfo {volume} {111}},\ \bibinfo
  {pages} {13004} (\bibinfo {year} {2014})}\BibitemShut {NoStop}%
\bibitem [{\citenamefont {Paulose}\ \emph {et~al.}(2015)\citenamefont
  {Paulose}, \citenamefont {Chen},\ and\ \citenamefont
  {Vitelli}}]{paulose2015topological}%
  \BibitemOpen
  \bibfield  {author} {\bibinfo {author} {\bibfnamefont {J.}~\bibnamefont
  {Paulose}}, \bibinfo {author} {\bibfnamefont {B.~G.}\ \bibnamefont {Chen}}, \
  and\ \bibinfo {author} {\bibfnamefont {V.}~\bibnamefont {Vitelli}},\ }\href
  {\doibase 10.1038/nphys3185} {\bibfield  {journal} {\bibinfo  {journal} {Nat.
  Phys.}\ }\textbf {\bibinfo {volume} {11}},\ \bibinfo {pages} {153} (\bibinfo
  {year} {2015})}\BibitemShut {NoStop}%
\bibitem [{\citenamefont {Huber}(2016)}]{huber2016topological}%
  \BibitemOpen
  \bibfield  {author} {\bibinfo {author} {\bibfnamefont {S.~D.}\ \bibnamefont
  {Huber}},\ }\href {\doibase 10.1038/nphys3801} {\bibfield  {journal}
  {\bibinfo  {journal} {Nat. Phys.}\ }\textbf {\bibinfo {volume} {12}},\
  \bibinfo {pages} {621} (\bibinfo {year} {2016})}\BibitemShut {NoStop}%
\bibitem [{\citenamefont {Rechtsman}\ \emph {et~al.}(2013)\citenamefont
  {Rechtsman}, \citenamefont {Zeuner}, \citenamefont {Plotnik}, \citenamefont
  {Lumer}, \citenamefont {Podolsky}, \citenamefont {Dreisow}, \citenamefont
  {Nolte}, \citenamefont {Segev},\ and\ \citenamefont
  {Szameit}}]{rechtsman2013photonic}%
  \BibitemOpen
  \bibfield  {author} {\bibinfo {author} {\bibfnamefont {M.~C.}\ \bibnamefont
  {Rechtsman}}, \bibinfo {author} {\bibfnamefont {J.~M.}\ \bibnamefont
  {Zeuner}}, \bibinfo {author} {\bibfnamefont {Y.}~\bibnamefont {Plotnik}},
  \bibinfo {author} {\bibfnamefont {Y.}~\bibnamefont {Lumer}}, \bibinfo
  {author} {\bibfnamefont {D.}~\bibnamefont {Podolsky}}, \bibinfo {author}
  {\bibfnamefont {F.}~\bibnamefont {Dreisow}}, \bibinfo {author} {\bibfnamefont
  {S.}~\bibnamefont {Nolte}}, \bibinfo {author} {\bibfnamefont
  {M.}~\bibnamefont {Segev}}, \ and\ \bibinfo {author} {\bibfnamefont
  {A.}~\bibnamefont {Szameit}},\ }\href {\doibase 10.1038/nature12066}
  {\bibfield  {journal} {\bibinfo  {journal} {Nature}\ }\textbf {\bibinfo
  {volume} {496}},\ \bibinfo {pages} {196} (\bibinfo {year}
  {2013})}\BibitemShut {NoStop}%
\bibitem [{\citenamefont {Lu}\ \emph {et~al.}(2016)\citenamefont {Lu},
  \citenamefont {Joannopoulos},\ and\ \citenamefont
  {Solja{\v{c}}i{\'c}}}]{lu2016topological}%
  \BibitemOpen
  \bibfield  {author} {\bibinfo {author} {\bibfnamefont {L.}~\bibnamefont
  {Lu}}, \bibinfo {author} {\bibfnamefont {J.~D.}\ \bibnamefont
  {Joannopoulos}}, \ and\ \bibinfo {author} {\bibfnamefont {M.}~\bibnamefont
  {Solja{\v{c}}i{\'c}}},\ }\href {\doibase 10.1038/nphys3796} {\bibfield
  {journal} {\bibinfo  {journal} {Nat. Phys.}\ }\textbf {\bibinfo {volume}
  {12}},\ \bibinfo {pages} {626} (\bibinfo {year} {2016})}\BibitemShut
  {NoStop}%
\bibitem [{\citenamefont {Agull\'o-Rueda}\ \emph {et~al.}(1989)\citenamefont
  {Agull\'o-Rueda}, \citenamefont {Mendez},\ and\ \citenamefont
  {Hong}}]{PhysRevB.40.1357}%
  \BibitemOpen
  \bibfield  {author} {\bibinfo {author} {\bibfnamefont {F.}~\bibnamefont
  {Agull\'o-Rueda}}, \bibinfo {author} {\bibfnamefont {E.~E.}\ \bibnamefont
  {Mendez}}, \ and\ \bibinfo {author} {\bibfnamefont {J.~M.}\ \bibnamefont
  {Hong}},\ }\href {\doibase 10.1103/PhysRevB.40.1357} {\bibfield  {journal}
  {\bibinfo  {journal} {Phys. Rev. B}\ }\textbf {\bibinfo {volume} {40}},\
  \bibinfo {pages} {1357} (\bibinfo {year} {1989})}\BibitemShut {NoStop}%
\bibitem [{\citenamefont {Kim}\ \emph {et~al.}(2011)\citenamefont {Kim},
  \citenamefont {Kusudo}, \citenamefont {Wu}, \citenamefont {Masumoto},
  \citenamefont {L{\"o}ffler}, \citenamefont {H{\"o}fling}, \citenamefont
  {Kumada}, \citenamefont {Worschech}, \citenamefont {Forchel},\ and\
  \citenamefont {Yamamoto}}]{kim2011dynamical}%
  \BibitemOpen
  \bibfield  {author} {\bibinfo {author} {\bibfnamefont {N.~Y.}\ \bibnamefont
  {Kim}}, \bibinfo {author} {\bibfnamefont {K.}~\bibnamefont {Kusudo}},
  \bibinfo {author} {\bibfnamefont {C.}~\bibnamefont {Wu}}, \bibinfo {author}
  {\bibfnamefont {N.}~\bibnamefont {Masumoto}}, \bibinfo {author}
  {\bibfnamefont {A.}~\bibnamefont {L{\"o}ffler}}, \bibinfo {author}
  {\bibfnamefont {S.}~\bibnamefont {H{\"o}fling}}, \bibinfo {author}
  {\bibfnamefont {N.}~\bibnamefont {Kumada}}, \bibinfo {author} {\bibfnamefont
  {L.}~\bibnamefont {Worschech}}, \bibinfo {author} {\bibfnamefont
  {A.}~\bibnamefont {Forchel}}, \ and\ \bibinfo {author} {\bibfnamefont
  {Y.}~\bibnamefont {Yamamoto}},\ }\href {\doibase 10.1038/nphys2012}
  {\bibfield  {journal} {\bibinfo  {journal} {Nat. Phys.}\ }\textbf {\bibinfo
  {volume} {7}},\ \bibinfo {pages} {681} (\bibinfo {year} {2011})}\BibitemShut
  {NoStop}%
\bibitem [{\citenamefont {Sun}\ \emph {et~al.}(2011)\citenamefont {Sun},
  \citenamefont {Gu}, \citenamefont {Katsura},\ and\ \citenamefont
  {Das~Sarma}}]{PhysRevLett.106.236803}%
  \BibitemOpen
  \bibfield  {author} {\bibinfo {author} {\bibfnamefont {K.}~\bibnamefont
  {Sun}}, \bibinfo {author} {\bibfnamefont {Z.}~\bibnamefont {Gu}}, \bibinfo
  {author} {\bibfnamefont {H.}~\bibnamefont {Katsura}}, \ and\ \bibinfo
  {author} {\bibfnamefont {S.}~\bibnamefont {Das~Sarma}},\ }\href {\doibase
  10.1103/PhysRevLett.106.236803} {\bibfield  {journal} {\bibinfo  {journal}
  {Phys. Rev. Lett.}\ }\textbf {\bibinfo {volume} {106}},\ \bibinfo {pages}
  {236803} (\bibinfo {year} {2011})}\BibitemShut {NoStop}%
\bibitem [{\citenamefont {Chern}\ \emph {et~al.}(2014)\citenamefont {Chern},
  \citenamefont {Chien},\ and\ \citenamefont {Di~Ventra}}]{PhysRevA.90.013609}%
  \BibitemOpen
  \bibfield  {author} {\bibinfo {author} {\bibfnamefont {G.-W.}\ \bibnamefont
  {Chern}}, \bibinfo {author} {\bibfnamefont {C.-C.}\ \bibnamefont {Chien}}, \
  and\ \bibinfo {author} {\bibfnamefont {M.}~\bibnamefont {Di~Ventra}},\ }\href
  {\doibase 10.1103/PhysRevA.90.013609} {\bibfield  {journal} {\bibinfo
  {journal} {Phys. Rev. A}\ }\textbf {\bibinfo {volume} {90}},\ \bibinfo
  {pages} {013609} (\bibinfo {year} {2014})}\BibitemShut {NoStop}%
\bibitem [{\citenamefont {Shen}(2013)}]{dirac}%
  \BibitemOpen
  \bibfield  {author} {\bibinfo {author} {\bibfnamefont {S.-Q.}\ \bibnamefont
  {Shen}},\ }\href {\doibase 10.1007/978-3-642-32858-9} {\emph {\bibinfo
  {title} {Topological insulators: Dirac equation in condensed matters}}}\
  (\bibinfo  {publisher} {Springer Science \& Business Media},\ \bibinfo {year}
  {2013})\BibitemShut {NoStop}%
\end{thebibliography}
\end{document}